\documentclass[10pt,twocolumn,letterpaper]{article}
\usepackage{iccv}
\usepackage{times}
\usepackage{epsfig}
\usepackage{graphicx}
\usepackage{amsmath}
\usepackage{amssymb}
\usepackage[breaklinks=true,bookmarks=false]{hyperref}

\iccvfinalcopy 
\def\iccvPaperID{9} 

\ificcvfinal\fi

\begin{document}

\title{Rapid Flood Inundation Forecast Using Fourier Neural Operator}

\author{Alexander Y. Sun$^1$$^*$
\and
Zhi Li$^2$ \\ 
\and
Wonhyun Lee$^1$ \\
\and 
Qixing Huang $^1$ \\
\and
Bridget R. Scanlon $^1$ \\
\and
Clint Dawson $^1$ \\
$^1$\text{The University of Texas at Austin, Austin, TX, USA} \\
$^2$\text{University of Oklahoma, Norman, OK, USA} \\
{\tt\small Corresponding author: alex.sun@beg.utexas.edu}
}

\maketitle
\ificcvfinal\thispagestyle{empty}\fi

\begin{abstract}
   Flood inundation forecast provides critical information for emergency planning before and during flood events. Real time flood inundation forecast tools are still lacking. High-resolution hydrodynamic modeling has become more accessible in recent years, however, predicting flood extents at the street and building levels in real-time is still computationally demanding. Here we present a hybrid process-based and data-driven machine learning (ML) approach for flood extent and inundation depth prediction. We used the Fourier neural operator (FNO), a highly efficient ML method, for surrogate modeling. The FNO model is demonstrated over an urban area in Houston (Texas, U.S.) by training using simulated water depths (in 15-min intervals) from six historical storm events and then tested over two holdout events. Results show FNO outperforms the baseline U-Net model. It maintains high predictability at all lead times tested (up to 3 hrs) and performs well when applying to new sites, suggesting strong generalization skill.   
\end{abstract}

\section{Introduction and application context}

Flooding is the most disruptive natural disaster, causing tens of billions of dollars of direct economic loss each year and affecting millions of people \cite{bates2022flood,merz2021causes}. In coastal areas, flooding may result from  overbank river flow (fluvial), heavy rainfall (pluvial), coastal storm surge, or a combination of all three. A warming climate is likely to further intensify the extreme precipitation, induce global sea level rise, and increase the frequency and intensity of tropical cyclones, making future flooding events more severe \cite{hirabayashi2013global, zscheischler2018future}. In the U.S., tens of millions of people are already exposed to the risk of coastal flooding \cite{sweet20222022}. By 2050, the U.S. population density in flood-prone coastal zones and megacities is expected to grow by 25\% \cite{aerts2014evaluating}, and flood risk is projected to increase by 26\%, with hotspots expected in highly populated counties along both coasts, as well as across the Northeast and through Appalachia \cite{bates2021combined,wing2022inequitable}.  

Flood inundation modeling (FIM), seeking to predict the flood water extent and depth using hydrodynamic models, is an integral part of flood risk management. Two major usages of FIM may be identified, flood susceptibility mapping and real time forecasting. In flood susceptibility mapping, FIM is used to quantify risks to flood events of a particular return period (e.g., 100-year event), providing risk-informed inputs to planners and insurers for land use zoning and infrastructure development. In real time forecasting, FIM is used to provide prediction of surface water levels during storm events. State of the art hydrodynamic models typically solve 2D full shallow water equations (SWE), which are simplified Navier–Stokes equations representing depth-averaged mass and momentum conservation \cite{bates2022flood}. A flood inundation model is forced by initial and boundary conditions such as upstream inflow and precipitation. For urban settings, the model spatial resolution should ideally be 3--10 m and the temporal resolution should be sub-hourly \cite{bates2022flood}. High-resolution FIM is not only necessary for street-level flood impact mapping, but also helps to analyze the exposure and vulnerability of local communities, especially disadvantaged population groups \cite{bates2023uneven, preisser2022intersecting, sanders2023large}. However, solving SWE at high spatiotemporal resolutions is still computationally  demanding, presenting a major challenge to its operational use. 

AI/ML-enabled surrogate models can provide a potential solution to scaling up FIM. In a broader context, AI/ML is envisioned to ultimately power the development of  earth system digital twins  \cite{bauer2021digital, karniadakis2021physics}. Extreme weather forecasting represents a major component of earth system digital twins. To demonstrate such a potential, this study presents a physics-based, hybrid ML approach for FIM. Physics-informed ML models are now widely developed and used in climate and earth system sciences to incorporate domain knowledge stemming from empirical and physical principles \cite{cuomo2022scientific,karniadakis2021physics,kashinath2021physics}. Integration of prior knowledge and/or physical constraints not only allows for training of more accurate ML models with sparse/noisy data, but also leads to more interpretable results. Hybrid ML, which is a form of physics-informed ML, utilizes outputs from process-based models as inputs to ML models. The ML algorithm we adopted in this work is Fourier Neural Operator (FNO), which is a type of neural operator for approximating mappings between infinite-dimensional function spaces \cite{li2020fourier}. FNO converts spatial domain representation into the spectral space through Fourier transforms, thus enabling more efficient computation of convolutions  \cite{kovachki2021neural}. 

\textbf{Main Contributions}. The main contributions of this work are summarized below: 
\begin{itemize}
    \item We developed an  FNO-based, flood inundation model for real-time flood mapping at multiple lead times 
    \item A physics-based loss function is used to minimize  the mismatch in predicted water depths and in spatial derivatives
    \item Results, demonstrated over an urban area in Houston, suggest the FNO model is more efficient than a U-Net-based baseline, and finally
    \item We showed that an FNO model pretrained on similar domains can be applied to the study site without fine-tuning, suggesting good  generalization capability 
\end{itemize}

\section{Related work}

FIM has been commonly used for flood susceptibility mapping \cite{bentivoglio2022deep}. Applications of (near) real time flood inundation mapping have only risen in recent years because of increased availability of SWE solvers and AI/ML. In \cite{zahura2020training}, a random forest model was trained to map topographic and environmental features to hourly water depths simulated by a hydrodynamic model at 16,914 street segments in the coastal city of Norfolk (Virginia, U.S.). Guo et al. \cite{guo2021data} presented a data-driven approach for maximum water depth prediction using CNN but assumed steady state. In \cite{hofmann2021floodgan}, a generative adversarial network (GAN)  was trained using synthetic rainfall events and simulated water depths. Their GAN-based approach was recently extended to include static features (e.g., elevation and slope) such that the trained model can be applied in zero-shot learning \cite{do2023generalizing}. Note that many of these previous FIM studies only considered a single lead time \cite{do2023generalizing,seleem2023transferability}. Satellite remote sensing provides flood extent information, but the coarse spatial resolution and latency of most satellites largely restrict their use to post-event impact assessment, such as mapping flooded areas using  multispectral surface reflectance imagery \cite{peng2019patch}, identifying the flood water extent from synthetic aperture radar (SAR) or multi-spectral (MS) imagery \cite{konapala2021exploring, shen2019inundation}. To the best of our knowledge, neural operators have not been applied to FIM.

\section{Methods}
\textbf{FNO} The goal of neural operator learning is to learn a mapping between two infinite-dimensional spaces by using paired inputs/outputs. Specifically, let $\mathcal{D}\in\mathbb{R}^d$ and $\mathcal{D'}\in\mathbb{R}^{d'}$ be bounded domains, and $\mathcal{A}$ and $\mathcal{U}$ be input and output function spaces defined on $\mathcal{D}$ and $\mathcal{D'}$, respectively. In our case, the input space consists of meteorological forcing (precipitation), static features (e.g., elevation) and/or antecedent water depths, while the output space represents the predicted surface water depths. Let $\mathcal{G}^{\dagger}$ represent an operator that maps the input to output, $\mathcal{G}^{\dagger}: \mathcal{A}\longrightarrow\mathcal{U}$. A neural operator $\mathcal{G}_{\theta}$ is a parametric map that approximates $\mathcal{G}^{\dagger}$, where $\theta\in\Theta$ are trainable parameters that can be obtained by solving a minimization problem with a loss function $L$ \cite{li2020fourier}
\begin{equation}
    \min_{\theta\in\Theta}\mathbb{E}\left(L(G_{\theta}(a), G^{\dagger}(a))\right),
\end{equation}
FNO seeks to approximate the following integral kernel operator commonly used in solutions of partial differential equations \cite{li2020fourier}
\begin{equation}
    \left(\mathcal{K}(v_l)\right)(x) = \int_D \kappa(x,y)v_l(y)dv_l(y),
\end{equation}
where $x,y\in\mathcal{D}$, $v_l$ is a function, and $\kappa(x,y)$ is a kernel function. In the Fourier integral operator, the kernel function is replaced by a convolution operator 
\begin{equation}
    \left(\mathcal{K}(v_l)\right)(x) = \mathcal{F}^{-1}\left(R_{\phi}\cdot\mathcal{F}(v_l)\right)(x),
\end{equation}
in which $\mathcal{F}$ and $\mathcal{F}^{-1}$ are forward and inverse Fourier transforms, and $R_{\phi}$ is the Fourier transform of a periodic function $\kappa$ that is parameterized by $\phi\in\Theta$. Assuming uniform discretization, then $\mathcal{F}$ is replaced by Fast Fourier Transform (FFT), and $R_{\phi}$ is approximated as a complex-valued tensor comprising a collection of truncated Fourier modes \cite{li2020fourier} and the values of $R_{\phi}$ are learned from training data.

\textbf{Loss function} We used the relative $L_2$ error as loss  function, which has been observed to impose a good normalization and regularization effect that prevents overfitting \cite{kovachki2021neural}. Inspired by physics informed ML,  we further minimized mismatch of spatial derivatives in terms of relative  $L_2$ error \cite{raissi2019physics,wen2022u} 
\begin{equation}
\begin{aligned}
    \mathrm{Rel. \, Loss} = \frac{\| u-\hat{u}\|}{\| u \|} + \beta_1\frac{\| du/dx-d\hat{u}/dx\|} {\| du/dx \|} \\ 
    + \beta_2\frac{\| du/dy-d\hat{u}/dy\|}{\| du/dy \|}  
\end{aligned}
\end{equation}
where $u$ and $\hat{u}$ are simulated and predicted water depths, $dx$, $\beta_1$ and $\beta_2$ are hyperparameters. We assigned $dx=0.2$ and used 0.1 for both $\beta_1$ and $\beta_2$ after a grid search.

\section{Experiments}
\begin{figure}[t]
\begin{center}
   \includegraphics[width=0.8\linewidth]{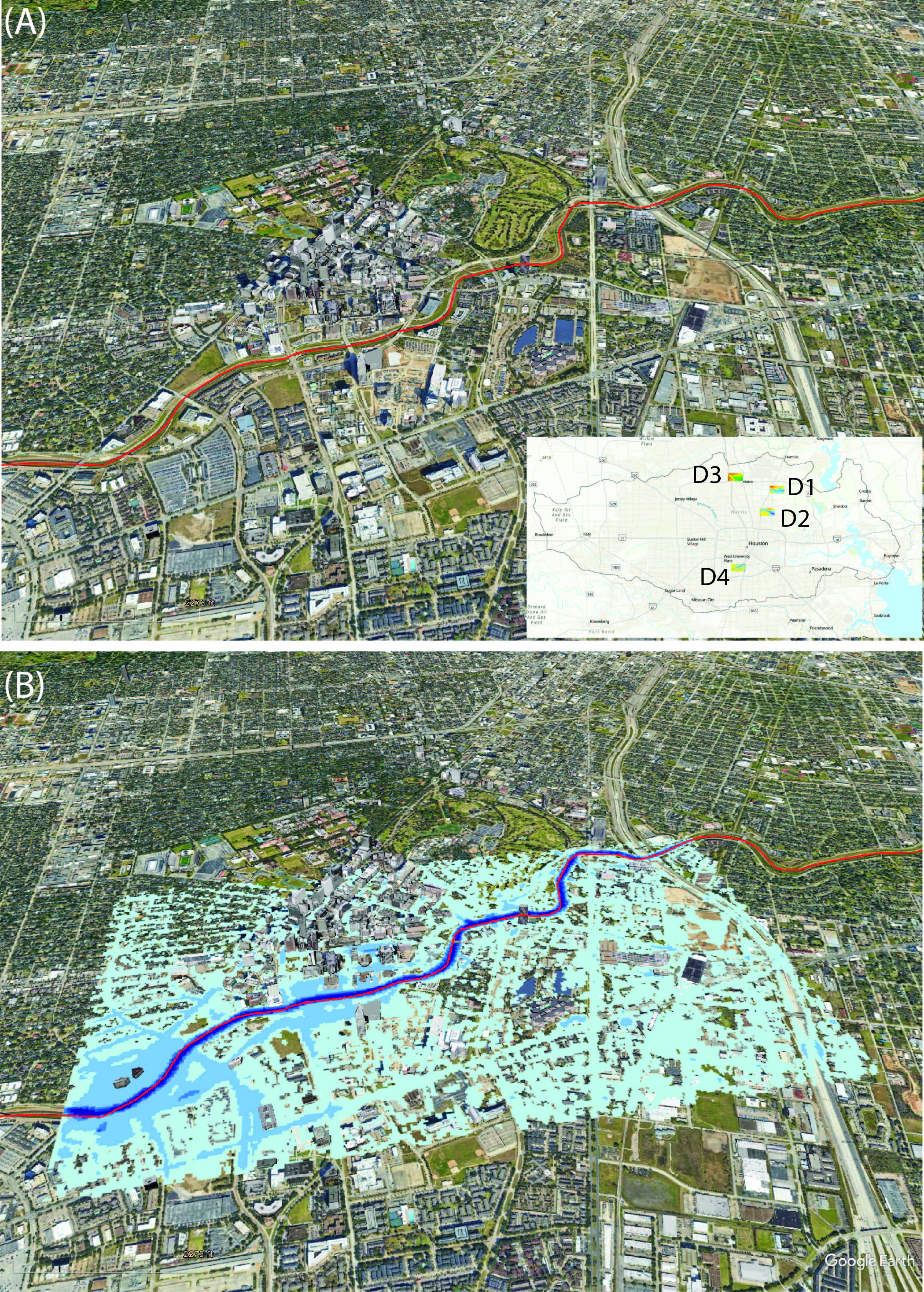}
\end{center}
   \caption{(A) Areal view of the Domain 4 (D4), which is located in the Brays Bayou watershed in Houston, Texas, U.S. and (B) an exemplary flooding scene, where darker blue indicates deeper water. Map inset shows locations of all domains (D1--D4) used in this study. }
\label{fig:1}
\end{figure}

\textbf{Physics model setup and dataset creation} The efficacy of FNO was demonstrated via a series of experiments. Under the \textit{single-domain, multi-event setting}, we considered a  $1.3\times2.6\mathrm{km}^2$  domain located in the Brays Bayou watershed near downtown Houston (D4 in Fig.\ref{fig:1}A). Brays Bayou is a fully urbanized watershed. Land use comprises of residential, industrial, and commercial buildings. The bayou flows eastward to its confluence with the Houston Ship Channel. Multiple historical storm events were simulated using the open-source CREST-iMAP,  a coupled hydrology-hydraulic framework for riverbank flow and overland flood inundation modeling \cite{li2021crest}. Previously, CREST-iMAP has been validated against Hurricane Harvey observations (e.g., streamflows, high water marks, and flood insurance claims) and showed comparable or better performance than other state-of-the-art hydrodynamic models \cite{li2021crest}. In this study, CREST-iMAP was forced by using a high quality, radar-based quantitative precipitation estimation (QPE) product---the Multi-Radar/Multi-Sensor System (MRMS) data published by the National Severe Storms Laboratory (NSSL) in U.S. National Oceanic and Atmospheric Administration (NOAA). MRMS comes at 1-km resolution in 2-min intervals. It is behind several rainfall nowcasting DL models such as Google's MetNet models \cite{espeholt2022deep,sonderby2020metnet}. In Fig.\ref{fig:1}B, a typical flooding scene is shown. We used simulations corresponding to six storm events from NOAA storm inventory for training and the rest for testing (Table\ref{tab:A1}. 
Under the \textit{multi-domain, multi-event setting}, the same NOAA storm events were first simulated over different spatial domains (D1--D3 in Fig.\ref{fig:1}A). We then trained an FNO model using D1--D3 data and tested on D4. More details on CREST-iMAP run configurations are provided under Appendix \ref{sec:a1}. 

\begin{figure}[t]
\begin{center}
   \includegraphics[width=0.95\linewidth]{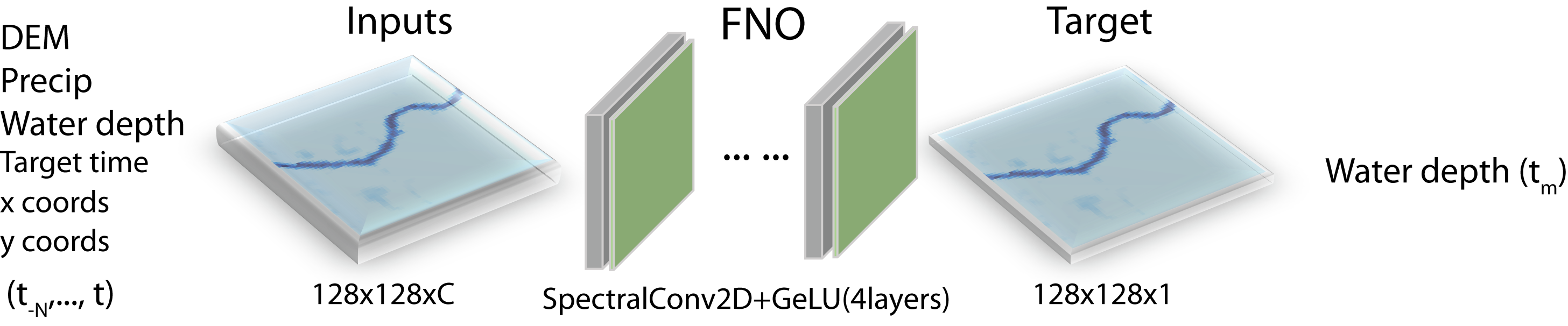}
\end{center}
   \caption{Architecture of FNO model, which consists of 4 blocks of spectral convolution layers (SpectralConv2d), each followed by a Gaussian Error Linear Units (GeLU) activation layer, where $N$ is the number lookback frames, $m$ is a random future frame, $C$ is the channel dimension representing the number of static frames (e.g., DEM) and dynamics frames.}
\label{fig:2}
\end{figure}

\textbf{Surrogate model training} A hybrid FNO surrogate model was trained for multi-step flood inundation prediction. We assumed the process-based flood inundation model and the FNO are running in parallel such that the outputs of the process-based model are available for the FNO model to ingest as predictors \cite{sun2022graph, slater2023hybrid}. To generate training datasets, we aggregated the CREST-iMAP inputs/outputs to 15-min intervals, and then sampled $128\times128$ input and output patches from the spatial domain. Each 15-min frame was sampled twice by randomly varying image centers within the image bound. Fig.\ref{fig:2} illustrates the architecture of the model. Inputs to FNO include antecedent precipitation and simulated water depths, and digital elevation model (DEM). A key feature of FNO is it concatenates x- and y-coordinates to the input features to help it capture dependencies between inputs and spatial locations, enabling the model to generalize to new locations \cite{li2020fourier}. The target variable is predicted water depth at a lead time, for which the target lead time value is concatenated to the inputs as a label. Alternatively, the target lead time can be treated as a  trainable feature so that the  trained model works for arbitrary lead times. The strategy was not used in this preliminary work. The FNO architecture includes four 2D spectral convolution layers, each followed by a GeLU activation layer. The data samples are split into training, validation, and testing in 0.8, 0.1, 0.1 ratios. The number of Fourier modes used is 16 in both directions. 

We trained the models in PyTorch Lightning \cite{falcon2019pytorch} using the Adam optimizer with an initial learning rate of 5e-4, a cosine annealing training schedule \cite{loshchilov2016sgdr}, and early stopping. Unless otherwise specified, the maximum epochs used is 60 and batch size is 8, which were found sufficient for this problem. Training was done on an Nvidia RTX3090 GPU. Training time per epoch is 72 sec wallclock time and inference time is 0.002 sec/sample.  For baseline, we considered a U-Net like model adapted from RainNet, which has a relatively simple deep architecture but nonetheless performs surprisingly well on radar-based precipitation nowcasting problems  \cite{ayzel2020rainnet}. RainNet still has 31.4M trainable parameters, while FNO has 8.1M. More details on the baseline model can be found in \ref{sec:a2}.

\textbf{Performance metrics} Model performance is measured using Critical Success Index (CSI, range 0--1.0) and mean absolute error (MAE, range 0--$\infty$) that are often used in FIM \cite{sonderby2020metnet,preisser2022intersecting}. CSI is defined as $\#\mathrm{Hits}/(\#\mathrm{Hits}+\#\mathrm{Misses}+\#\mathrm{FalseAlarms})$ , where \#Hits is the number of flood events correctly predicted; \#Misses is the number of  flood events incorrectly predicted as non-flood events; and \#FalseAlarms is the number of non-flood events incorrectly predicted as flood events \cite{clark2014conus}. In this work, we used the average of CSI calculated over three depth thresholds, 3cm, 10cm, and 25cm, to gauge model performance. In the literature 3cm is often used as the threshold for nuance flooding \cite{moftakhari2018nuisance}. Both CSI and MAE were calculated at the grid cell level and then averaged spatially and temporally. 

\section{Results}
\begin{figure}[t]
\begin{center}
\includegraphics[width=0.95\linewidth]{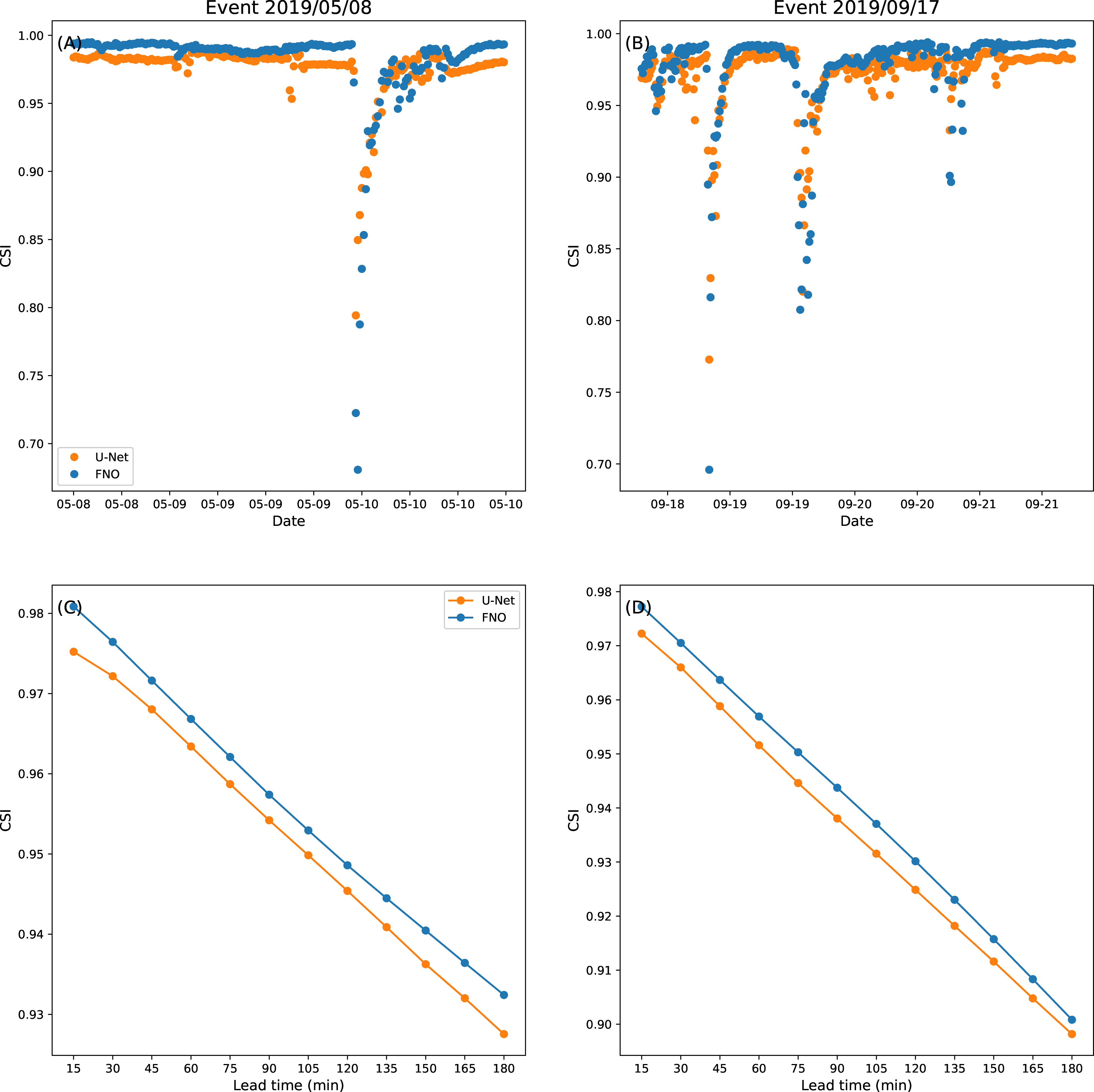}
\end{center}
   \caption{Comparison between U-Net and FNO: spatially averaged critical success index (CSI) at the 15-min lead time (A, B), and spatiotemporally averaged CSI at all lead times (C, D) for two test events.}
\label{fig:3}
\end{figure}

\textbf{Single-domain multi-event results} Table\ref{tab:2} summarizes CSI and MAE metrics on two holdout storm events. Note the event on 2019/09/17 corresponds to Tropical Storm Imelda, which is the fourth wettest event on record in Texas \cite{smith20202010}. At 12 lookback frames (i.e., 180min), FNO-12 outperformed U-Net-12 for all 12 lead times (Fig.\ref{fig:3}). The CSI of both models drops at the beginning of streamflow ascending due to strong discontinuity in predictors, but quickly bounces back for the rest of flood duration. As lead time increases, the FNO performance decreases linearly; thus, it is expected to give reasonable performance for longer lead times than tested here. 

As ablation studies, we considered longer lookback periods (24 past frames or 6 hrs), which did not improve the results (FNO-24 in Table\ref{tab:2}). This is probably because of the lack of long memory during storm events. We also considered a precipitation-only experiment (FNO-12P in Table\ref{tab:2}) where no antecedent water depth information is used. In that case the CSI dropped significantly, suggesting the importance of hybrid forecasting. 

\textbf{Multi-domain multi-event results} An FNO model pretrained using data from D1--D3 was applied to predict the same events and results for D4. Results, shown in Table \ref{tab:2} under FNO-MD, suggest the pretrained model adapts to the new domain well in this case, largely because of the physics embedded.

\begin{table}
\caption{Metrics obtained for 15min and 180min lead times for two test storm events, MAE---mean absolute error, CSI---critical success index,  numbers at the end of model names indicate the length of lookback period, and MD indicates multdomain}
\begin{center}
\begin{tabular}{l c c c}
\hline
Event & Model & 15min & 180min \\
      &       &  (CSI, MAE)   & (CSI, MAE) \\
\hline
2019/05/08 & U-Net-12    & 0.9752, 0.0056 & 0.9275, 0.0168 \\
           & FNO-12      & \textbf{0.9808}, \textbf{0.0044} & \textbf{0.9324}, \textbf{0.0153} \\
           & FNO-24      & 0.9764, 0.0048 & 0.9240, 0.0193 \\          
           & FNO-12P     & 0.4861, 0.0825 & 0.4855, 0.0822 \\ 
           & FNO-MD      & 0.9659, 0.0083 & 0.9032, 0.0218 \\
\hline           
2019/09/17 & U-Net-12 & 0.9723, 0.0088 & 0.8982, 0.0320  \\  
           & FNO-12   & 0.9773, 0.0075 & 0.9008, 0.0286  \\  
           & FNO-24   & 0.9683, 0.0099 & 0.8914, 0.0348  \\
           & FNO-12P  & 0.4550, 0.1197 & 0.4470, 0.1187  \\  
           & FNO-MD & 0.9646, 0.0139 & 0.8895,0.0400 \\
\hline
\end{tabular}
\end{center}
\label{tab:2}
\end{table}

\section{Conclusion}
We have developed a Fourier Neural Operator (FNO) model for multi lead time rapid flood inundation modeling (FIM) by combining process-based modeling with data-driven ML. Results indicate the hybrid FNO learns the input/output mappings of the underlying hydrodynamic model well. The model may be deployed in parallel with the process-based modeling for rapid FIM.   
\section*{Acknowledgement} AYS, WL, BRS, and CW were funded by Department of Energy Advanced Scientific Computing Research program under grant no. DE-SC0022211.

\section*{Appendix}
\setcounter{table}{0}
\renewcommand{\thetable}{A\arabic{table}}
\setcounter{figure}{0}
\renewcommand{\thefigure}{A\arabic{figure}}
\setcounter{section}{0}
\renewcommand{\thesection}{A\arabic{section}}

\section{Model configuration}\label{sec:a1}
\begin{figure}[t]
\begin{center}
   \includegraphics[width=0.8\linewidth]{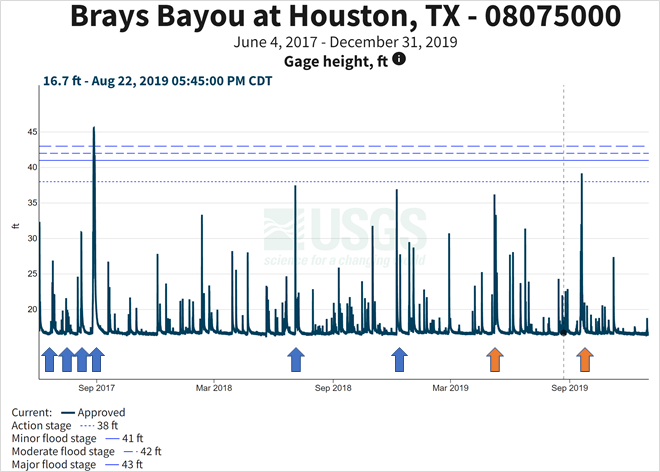}
\end{center}
   \caption{Observed hydrograph at USGS Gage No. 08075000 provides the upstream boundary condition for the Brays Bayou flood inundation  simulation. In the plot, blue color arrows indicate storm events used for training, while orange color arrows indciate storm events used for test.}
\label{fig:a1}
\end{figure}

Storm events were retrieved from U.S. National Oceanic and Atmospheric Administration (NOAA) catalog. Table \ref{tab:A1} shows the start and end dates of all storm events. The events are further visualized on the streamflow hydrograph in Fig. \ref{fig:a1}. 

Resolution of the digital elevation model (DEM) used in this study is 10m, which is obtained from U.S. Geological Survey (USGS) \url{https://www.usgs.gov/3d-elevation-program/about-3dep-products-services}. Streamflow (USGS Gage 08075000) is used as upper boundary condition for hydrodynamic model CREST-iMAP (Fig. \ref{fig:a1}. We used a radar-based precipitation forcing dataset, the MRMS data from NOAA \url{https://mrms.nssl.noaa.gov}. MRMS comes at spatial resolution of 4 km and 2-min temporal resolution. 

CREST-iMAP uses hydrological CREST to solve for water balance equation and ANUGA-Hydro for hydrulic modeling \cite{li2021crest}. 

\begin{table}
\caption{Storm events}
\begin{center}
\begin{tabular}{|l|c|c|}
\hline
Start Date & End Date & Usage \\
\hline\hline
2017/06/04 & 2017/06/07 & Training/validation \\
2017/06/24 & 2017/06/26 & Training/validation \\
2017/07/09 & 2017/07/13 & Training/validation \\
2017/08/26 & 2017/09/02 & Training/validation \\
2018/07/04 & 2018/07/07 & Training/validation \\
2018/12/07 & 2018/12/10 & Training/validation \\
2019/05/07 & 2019/05/11 & Testing \\
2019/09/19 & 2019/09/22 & Testing \\
\hline
\end{tabular}
\end{center}
\label{tab:A1}
\end{table}

\begin{figure}[t]
\begin{center}
   \includegraphics[width=0.8\linewidth]{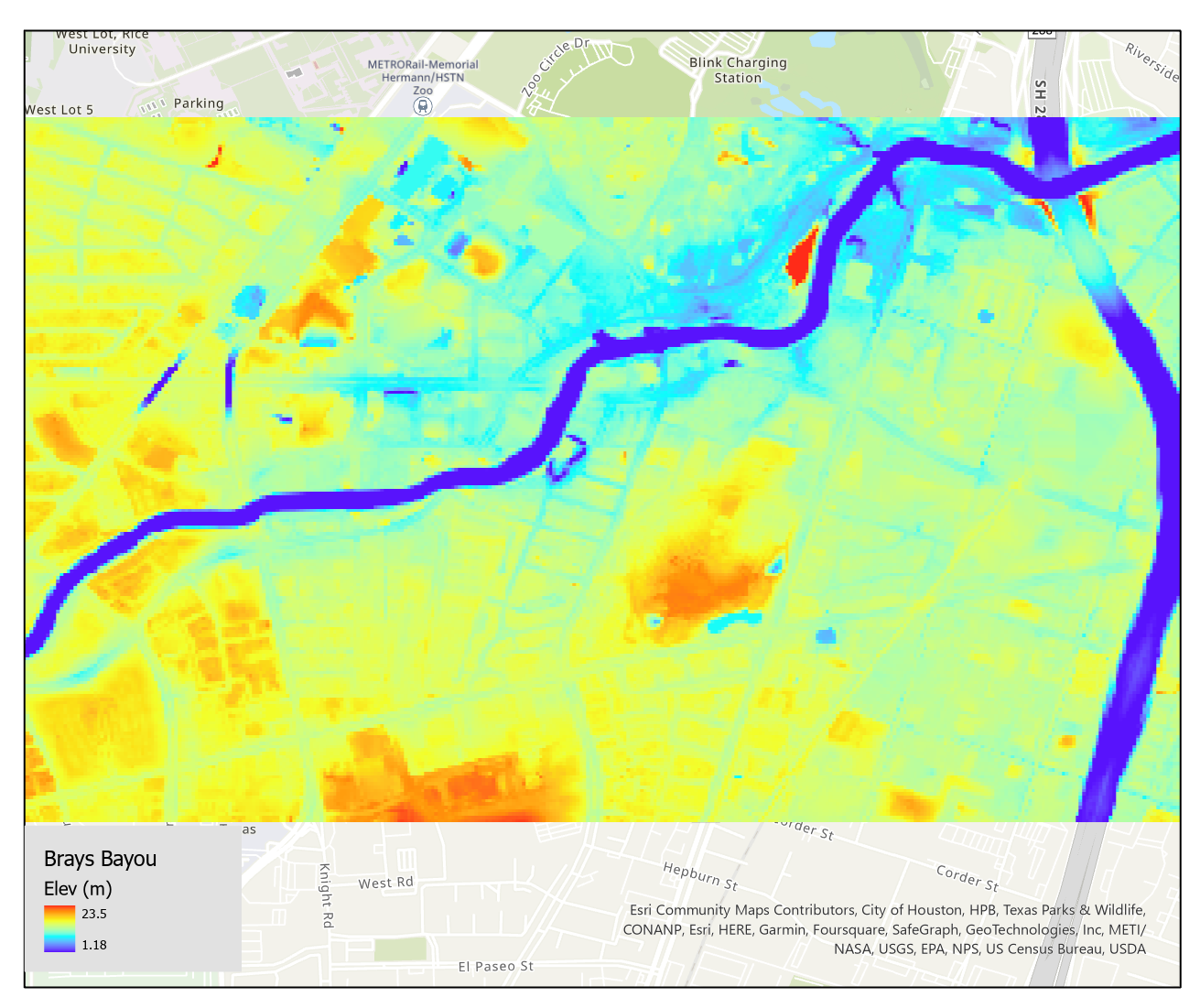}
\end{center}
   \caption{Digital elevation map (DEM) of the Brays Bayou study domain}
\label{fig:a2}
\end{figure}

\section{U-Net model architecture}\label{sec:a2}
The U-Net architecture is based on the PyTorch implementation of RainNet \cite{ayzel2020rainnet}, \url{https://github.com/fmidev/rainnet}.

\begin{table}
\caption{The U-Net model architecture is based on \cite{ayzel2020rainnet}. $C=d*n_{\textrm{lookback}}+s$ is the number of inputs,  where $d$ is number of dynamic variables, $n_{\textrm{lookback}}$ is the lookback length, and $s$ is number of static variables.}
\begin{center}
\begin{tabular}{|l|c|c|c|}
\hline
Layer & Feature In & Feature Out & kernel size \\   
\hline\hline
    Input  & C & 64 & 3 \\ 
    D1     & 64 & 128 & 3 \\
    Maxpool &  & & 2 \\ 
    D2     & 128 & 256 & 3 \\
    Maxpool &  & & 2 \\
    D3     & 256 & 512 & 3 \\
    Maxpool &  & & 2 \\
    D4     & 512 & 1024 & 3 \\
    Maxpool &  & & 2 \\
    U1     & 1536 & 512 & 3 \\
    Upsample & & &  2 \\
    U2     & 768 & 256 & 3 \\
    Upsample & & &  2 \\
    U3     & 384 & 128 & 3 \\
    Upsample & & &  2  \\
    U4     & 192 & 64 & 3 \\
    Upsample & & &  2  \\
    Last   & 64  & 2  & 3 \\
    Output & 2   & 1  & 1 \\
\hline
\end{tabular}
\end{center}
\label{tab:A2}
\end{table}

The RainNet model was trained using the log hyperbolic cosine loss function ($\mathrm{logcosh}$), which has been shown to be beneficial to precipitation nowcasting ML applications \cite{ayzel2020rainnet}
\begin{equation}
    \mathrm{Logcosh} = \frac{\sum_{i=1}^{n}\log(\cosh(\hat{u}_i - u_i))}{n},
\end{equation}
where $\cosh(x)=\frac{1}{2}(e^x + e^{-x})$, $u$ and $\hat{u}$ are "true" and predicted water depths.


\begin{thebibliography}{10}\itemsep=-1pt

\bibitem{aerts2014evaluating}
Jeroen~CJH Aerts, WJ~Wouter Botzen, Kerry Emanuel, Ning Lin, Hans De~Moel, and
  Erwann~O Michel-Kerjan.
\newblock Evaluating flood resilience strategies for coastal megacities.
\newblock {\em Science}, 344(6183):473--475, 2014.

\bibitem{ayzel2020rainnet}
Georgy Ayzel, Tobias Scheffer, and Maik Heistermann.
\newblock Rainnet v1. 0: a convolutional neural network for radar-based
  precipitation nowcasting.
\newblock {\em Geoscientific Model Development}, 13(6):2631--2644, 2020.

\bibitem{bates2023uneven}
Paul Bates.
\newblock Uneven burden of urban flooding.
\newblock {\em Nature Sustainability}, 6(1):9--10, 2023.

\bibitem{bates2022flood}
Paul~D Bates.
\newblock Flood inundation prediction.
\newblock {\em Annual Review of Fluid Mechanics}, 54:287--315, 2022.

\bibitem{bates2021combined}
Paul~D Bates, Niall Quinn, Christopher Sampson, Andrew Smith, Oliver Wing,
  Jeison Sosa, James Savage, Gaia Olcese, Jeff Neal, Guy Schumann, et~al.
\newblock Combined modeling of us fluvial, pluvial, and coastal flood hazard
  under current and future climates.
\newblock {\em Water Resources Research}, 57(2):e2020WR028673, 2021.

\bibitem{bauer2021digital}
Peter Bauer, Peter~D Dueben, Torsten Hoefler, Tiago Quintino, Thomas~C
  Schulthess, and Nils~P Wedi.
\newblock The digital revolution of earth-system science.
\newblock {\em Nature Computational Science}, 1(2):104--113, 2021.

\bibitem{bentivoglio2022deep}
Roberto Bentivoglio, Elvin Isufi, Sebastian~Nicolaas Jonkman, and Riccardo
  Taormina.
\newblock Deep learning methods for flood mapping: a review of existing
  applications and future research directions.
\newblock {\em Hydrology and Earth System Sciences}, 26(16):4345--4378, 2022.

\bibitem{clark2014conus}
Robert~A Clark, Jonathan~J Gourley, Zachary~L Flamig, Yang Hong, and Edward
  Clark.
\newblock Conus-wide evaluation of national weather service flash flood
  guidance products.
\newblock {\em Weather and Forecasting}, 29(2):377--392, 2014.

\bibitem{cuomo2022scientific}
Salvatore Cuomo, Vincenzo~Schiano Di~Cola, Fabio Giampaolo, Gianluigi Rozza,
  Maziar Raissi, and Francesco Piccialli.
\newblock Scientific machine learning through physics--informed neural
  networks: where we are and what’s next.
\newblock {\em Journal of Scientific Computing}, 92(3):88, 2022.

\bibitem{do2023generalizing}
Cesar~AF do Lago, Marcio~H Giacomoni, Roberto Bentivoglio, Riccardo Taormina,
  Marcus N~Gomes Junior, and Eduardo~M Mendiondo.
\newblock Generalizing rapid flood predictions to unseen urban catchments with
  conditional generative adversarial networks.
\newblock {\em Journal of Hydrology}, 618:129276, 2023.

\bibitem{espeholt2022deep}
Lasse Espeholt, Shreya Agrawal, Casper S{\o}nderby, Manoj Kumar, Jonathan Heek,
  Carla Bromberg, Cenk Gazen, Rob Carver, Marcin Andrychowicz, Jason Hickey,
  et~al.
\newblock Deep learning for twelve hour precipitation forecasts.
\newblock {\em Nature communications}, 13(1):5145, 2022.

\bibitem{falcon2019pytorch}
William~A Falcon.
\newblock Pytorch lightning.
\newblock {\em GitHub}, 3, 2019.

\bibitem{guo2021data}
Zifeng Guo, Joao~P Leitao, Nuno~E Sim{\~o}es, and Vahid Moosavi.
\newblock Data-driven flood emulation: Speeding up urban flood predictions by
  deep convolutional neural networks.
\newblock {\em Journal of Flood Risk Management}, 14(1):e12684, 2021.

\bibitem{hirabayashi2013global}
Yukiko Hirabayashi, Roobavannan Mahendran, Sujan Koirala, Lisako Konoshima, Dai
  Yamazaki, Satoshi Watanabe, Hyungjun Kim, and Shinjiro Kanae.
\newblock Global flood risk under climate change.
\newblock {\em Nature climate change}, 3(9):816--821, 2013.

\bibitem{hofmann2021floodgan}
Julian Hofmann and Holger Sch{\"u}ttrumpf.
\newblock floodgan: Using deep adversarial learning to predict pluvial flooding
  in real time.
\newblock {\em Water}, 13(16):2255, 2021.

\bibitem{karniadakis2021physics}
George~Em Karniadakis, Ioannis~G Kevrekidis, Lu Lu, Paris Perdikaris, Sifan
  Wang, and Liu Yang.
\newblock Physics-informed machine learning.
\newblock {\em Nature Reviews Physics}, 3(6):422--440, 2021.

\bibitem{kashinath2021physics}
Karthik Kashinath, M Mustafa, Adrian Albert, JL Wu, C Jiang, Soheil
  Esmaeilzadeh, Kamyar Azizzadenesheli, R Wang, A Chattopadhyay, A Singh,
  et~al.
\newblock Physics-informed machine learning: case studies for weather and
  climate modelling.
\newblock {\em Philosophical Transactions of the Royal Society A},
  379(2194):20200093, 2021.

\bibitem{konapala2021exploring}
Goutam Konapala, Sujay~V Kumar, and Shahryar~Khalique Ahmad.
\newblock Exploring sentinel-1 and sentinel-2 diversity for flood inundation
  mapping using deep learning.
\newblock {\em ISPRS Journal of Photogrammetry and Remote Sensing},
  180:163--173, 2021.

\bibitem{kovachki2021neural}
Nikola Kovachki, Zongyi Li, Burigede Liu, Kamyar Azizzadenesheli, Kaushik
  Bhattacharya, Andrew Stuart, and Anima Anandkumar.
\newblock Neural operator: Learning maps between function spaces.
\newblock {\em arXiv preprint arXiv:2108.08481}, 2021.

\bibitem{li2021crest}
Zhi Li, Mengye Chen, Shang Gao, Xiangyu Luo, Jonathan~J Gourley, Pierre
  Kirstetter, Tiantian Yang, Randall Kolar, Amy McGovern, Yixin Wen, et~al.
\newblock Crest-imap v1. 0: A fully coupled hydrologic-hydraulic modeling
  framework dedicated to flood inundation mapping and prediction.
\newblock {\em Environmental Modelling \& Software}, 141:105051, 2021.

\bibitem{li2020fourier}
Zongyi Li, Nikola Kovachki, Kamyar Azizzadenesheli, Burigede Liu, Kaushik
  Bhattacharya, Andrew Stuart, and Anima Anandkumar.
\newblock Fourier neural operator for parametric partial differential
  equations.
\newblock {\em arXiv preprint arXiv:2010.08895}, 2020.

\bibitem{loshchilov2016sgdr}
Ilya Loshchilov and Frank Hutter.
\newblock Sgdr: Stochastic gradient descent with warm restarts.
\newblock {\em arXiv preprint arXiv:1608.03983}, 2016.

\bibitem{merz2021causes}
Bruno Merz, G{\"u}nter Bl{\"o}schl, Sergiy Vorogushyn, Francesco Dottori,
  Jeroen~CJH Aerts, Paul Bates, Miriam Bertola, Matthias Kemter, Heidi
  Kreibich, Upmanu Lall, et~al.
\newblock Causes, impacts and patterns of disastrous river floods.
\newblock {\em Nature Reviews Earth \& Environment}, 2(9):592--609, 2021.

\bibitem{moftakhari2018nuisance}
Hamed~R Moftakhari, Amir AghaKouchak, Brett~F Sanders, Maura Allaire, and
  Richard~A Matthew.
\newblock What is nuisance flooding? defining and monitoring an emerging
  challenge.
\newblock {\em Water Resources Research}, 54(7):4218--4227, 2018.

\bibitem{peng2019patch}
Bo Peng, Zonglin Meng, Qunying Huang, and Caixia Wang.
\newblock Patch similarity convolutional neural network for urban flood extent
  mapping using bi-temporal satellite multispectral imagery.
\newblock {\em Remote Sensing}, 11(21):2492, 2019.

\bibitem{preisser2022intersecting}
Matthew Preisser, Paola Passalacqua, R~Patrick Bixler, and Julian Hofmann.
\newblock Intersecting near-real time fluvial and pluvial inundation<?
  xmltex$\backslash$break?> estimates with sociodemographic vulnerability<?
  xmltex$\backslash$break?> to quantify a household flood impact index.
\newblock {\em Hydrology and Earth System Sciences}, 26(15):3941--3964, 2022.

\bibitem{raissi2019physics}
Maziar Raissi, Paris Perdikaris, and George~E Karniadakis.
\newblock Physics-informed neural networks: A deep learning framework for
  solving forward and inverse problems involving nonlinear partial differential
  equations.
\newblock {\em Journal of Computational physics}, 378:686--707, 2019.

\bibitem{sanders2023large}
Brett~F Sanders, Jochen~E Schubert, Daniel~T Kahl, Katharine~J Mach, David
  Brady, Amir AghaKouchak, Fonna Forman, Richard~A Matthew, Nicola Ulibarri,
  and Steven~J Davis.
\newblock Large and inequitable flood risks in los angeles, california.
\newblock {\em Nature sustainability}, 6(1):47--57, 2023.

\bibitem{seleem2023transferability}
Omar Seleem, Georgy Ayzel, Axel Bronstert, and Maik Heistermann.
\newblock Transferability of data-driven models to predict urban pluvial flood
  water depth in berlin, germany.
\newblock {\em Natural Hazards and Earth System Sciences}, 23(2):809--822,
  2023.

\bibitem{shen2019inundation}
Xinyi Shen, Dacheng Wang, Kebiao Mao, Emmanouil Anagnostou, and Yang Hong.
\newblock Inundation extent mapping by synthetic aperture radar: A review.
\newblock {\em Remote Sensing}, 11(7):879, 2019.

\bibitem{slater2023hybrid}
Louise~J Slater, Louise Arnal, Marie-Am{\'e}lie Boucher, Annie Y-Y Chang, Simon
  Moulds, Conor Murphy, Grey Nearing, Guy Shalev, Chaopeng Shen, Linda Speight,
  et~al.
\newblock Hybrid forecasting: blending climate predictions with ai models.
\newblock {\em Hydrology and earth system sciences}, 27(9):1865--1889, 2023.

\bibitem{smith20202010}
Adam Smith.
\newblock 2010--2019: A landmark decade of us. billion-dollar weather and
  climate disasters.
\newblock {\em National Oceanic and Atmospheric Administration}, 2020.

\bibitem{sonderby2020metnet}
Casper~Kaae S{\o}nderby, Lasse Espeholt, Jonathan Heek, Mostafa Dehghani,
  Avital Oliver, Tim Salimans, Shreya Agrawal, Jason Hickey, and Nal
  Kalchbrenner.
\newblock Metnet: A neural weather model for precipitation forecasting.
\newblock {\em arXiv preprint arXiv:2003.12140}, 2020.

\bibitem{sun2022graph}
Alexander~Y Sun, Peishi Jiang, Zong-Liang Yang, Yangxinyu Xie, and Xingyuan
  Chen.
\newblock A graph neural network (gnn) approach to basin-scale river network
  learning: the role of physics-based connectivity and data fusion.
\newblock {\em Hydrology and Earth System Sciences}, 26(19):5163--5184, 2022.

\bibitem{sweet20222022}
WV Sweet et~al.
\newblock 2022 sea level rise technical report.
\newblock 2022.

\bibitem{wen2022u}
Gege Wen, Zongyi Li, Kamyar Azizzadenesheli, Anima Anandkumar, and Sally~M
  Benson.
\newblock U-fno—an enhanced fourier neural operator-based deep-learning model
  for multiphase flow.
\newblock {\em Advances in Water Resources}, 163:104180, 2022.

\bibitem{wing2022inequitable}
Oliver~EJ Wing, William Lehman, Paul~D Bates, Christopher~C Sampson, Niall
  Quinn, Andrew~M Smith, Jeffrey~C Neal, Jeremy~R Porter, and Carolyn Kousky.
\newblock Inequitable patterns of us flood risk in the anthropocene.
\newblock {\em Nature Climate Change}, 12(2):156--162, 2022.

\bibitem{zahura2020training}
Faria~T Zahura, Jonathan~L Goodall, Jeffrey~M Sadler, Yawen Shen, Mohamed~M
  Morsy, and Madhur Behl.
\newblock Training machine learning surrogate models from a high-fidelity
  physics-based model: Application for real-time street-scale flood prediction
  in an urban coastal community.
\newblock {\em Water Resources Research}, 56(10):e2019WR027038, 2020.

\bibitem{zscheischler2018future}
Jakob Zscheischler, Seth Westra, Bart~JJM Van Den~Hurk, Sonia~I Seneviratne,
  Philip~J Ward, Andy Pitman, Amir AghaKouchak, David~N Bresch, Michael
  Leonard, Thomas Wahl, et~al.
\newblock Future climate risk from compound events.
\newblock {\em Nature Climate Change}, 8(6):469--477, 2018.

\end{thebibliography}
\end{document}